\documentclass[fleqn,10pt]{wlscirep}
\title{Inelastic electron injection in a water chain}

\author[1,*]{Valerio Rizzi}
\author[1]{Tchavdar N. Todorov}
\author[1]{Jorge J. Kohanoff}

\affil[1]{Atomistic Simulation Centre, Queen's University Belfast, Belfast BT7 1NN, Northern Ireland, United Kingdom}
\affil[*]{vrizzi01@qub.ac.uk}

\usepackage{epstopdf}
\usepackage{dcolumn}
\usepackage{bm}
\usepackage[mathlines]{lineno}
\usepackage[T1]{fontenc}
\usepackage{amsmath,amsfonts,amsthm,amssymb}
\usepackage[sc,osf]{mathpazo}   
\usepackage{siunitx}
\usepackage{listings,slashed,booktabs,braket}
\usepackage{color}
\usepackage{framed}
\usepackage{fancyhdr} 
\usepackage{multirow}
\usepackage[normalem]{ulem} 

\graphicspath{{figure/}}

\begin{abstract}
Irradiation of biological matter triggers a cascade of secondary particles that interact with their surroundings, resulting in damage. 
Low-energy electrons are one of the main secondary species and electron-phonon interaction plays a fundamental role in their dynamics.
We have developed a method to capture the electron-phonon inelastic energy exchange in real time and have used it to inject electrons into a simple system that models a biological environment, a water chain.
We simulated both an incoming electron pulse and a steady stream of electrons and found that electrons with energies just outside bands of excited molecular states can enter the chain through phonon emission or absorption. Furthermore, this phonon-assisted dynamical behaviour shows great sensitivity to the vibrational temperature, highlighting a crucial controlling factor for the injection and propagation of electrons in water.
\end{abstract}
\begin{document}
	
	\flushbottom
	\maketitle
	\thispagestyle{empty}
	
	\section*{Introduction}

When high-energy radiation penetrates living cells, it ionizes molecules along its path and can cause cell death by damaging DNA. 
Radiation events involve a sequence of processes that ultimately require a clear microscopic understanding \cite{Baccarelli2011}.
Only about one third of the cellular damage is produced by direct interaction of the ionizing radiation with DNA, while the rest is due to secondary species, produced in the first hundreds to thousands of femtoseconds following the primary irradiation of the system \cite{Michael2000}.

Secondary electrons are a key irradiation by-product, as about $\simeq 50 \cdot 10^3$ electrons are emitted for every MeV of incoming energy \cite{Alizadeh2015}. The majority of secondary electrons are low energy electrons (LEEs), with an energy distribution peaking below 10 eV \cite{Pimblott2007}.
It may seem intuitive that the higher the electronic energy, the more significant the damage, and that electrons with energies below the DNA ionization threshold $\approx 15$ eV cannot cause strand breaks and destroy DNA, but this perception was challenged in 2000 \cite{Boudaiffa2000,Michael2000}. 
It was discovered that LEEs with energies between 3 and 20 eV can damage DNA considerably and their damaging power does not constantly increase with their energy. 
These studies triggered an intense effort into understanding the interaction mechanisms of LEEs with DNA \cite{Martin2004,Cho2004,Simons2006,Baccarelli2011,Alizadeh2015}. 

The study of LEEs dynamics and their interaction with the cellular environment hinges on the non-adiabatic evolution of molecular systems over picosecond timescales. The mesoscale nature of this problem makes it especially challenging 
for non-adiabatic quantum electron-nuclear simulations.
We have recently reported a method, called Effective Correlated Electron-Ion Dynamics (ECEID), designed for such simulations \cite{Rizzi2016}. 
In this paper we combine ECEID with electronic Open Boundaries (OB) \cite{McEniry2007} to simulate in real time the injection of LEEs and their inelastic dynamical interaction with phonons.

Water is the main component of cells: most LEEs are generated from it and, in turn, interact with it, while its presence plays an enhancing role in DNA radiation damage \cite{Cooper2000}.
Electron tunneling through static water configurations has been studied \cite{Peskin1999,Galperin2001,Galperin2001a}, focusing on resonance lifetimes and the relevance of inelastic effects at a perturbative level.
We adopt a simple model of a water chain with one phonon mode to mimic a minimal biological environment and inject LEEs into it at different energies. Phonon absorption and emission play an essential role for enabling electrons to enter the water chain. 
Phonon-assisted injection shows a great sensitivity to vibrational temperature, with
the possibility of dramatically reducing or enhancing the electron flow.
Phonons are therefore a crucial control factor for the injection of LEEs into water.
The excited states display an energy dependent lifetime that we compare with self-energy results and offer a possible electron trapping mechanism.

It is hoped that these results will provide a framework for further dynamical simulations of radiation damage in more complex biological systems \cite{Smyth2011,McAllister2015}.

\section*{Theoretical Background}

ECEID is a real-time non-adiabatic electron-phonon method. Its full derivation is given in \cite{Rizzi2016} and is briefly summarised below. 
We start from Hamiltonian
\begin{equation}
\hat{H} = \underbrace{\hat{H}_{\mathrm{e}} + \sum_{\nu=1}^{N_{\mathrm{p}}} \Big( 
	\frac{\hat{P}_\nu^2}{2 M_\nu} + \frac{1}{2} K_\nu \hat{X}^2_\nu \Big)}_{\hat{H}_0} 
- \sum_{\nu=1}^{N_{\mathrm{p}}} \hat{F}_\nu \hat{X}_\nu .
\label{eq:hamone}
\end{equation}
where $\hat{H}_{\mathrm{0}}$ is the unperturbed Hamiltonian describing non-interacting phonons and electrons, with $\hat{H}_{\mathrm{e}}$ a general many-electron Hamiltonian. The $N_{\rm p}$ phonon modes are treated harmonically and each mode $\nu$ is coupled linearly to the electrons by the electronic operator $\hat{F}_\nu$. $M_\nu$ is a mass-like parameter, $K_\nu$ is a spring constant, $\omega_\nu = \sqrt{K_\nu/M_\nu}$ is an angular frequency, $\hat{X}_\nu$ and and $\hat{P}_\nu$ are the canonical displacement and momentum operators for that degree of freedom. In this work we will employ only one phonon, so, from now on, we suppress the subscript $\nu$. Notice that $\hat{F}$ and $\hat{H}$ are related through $\hat{F} = -\partial \hat{H} / \partial X$, enabling $\hat{F}$ to be constructed directly from a chosen Hamiltonian and a chosen generalized coordinate $X$. 

By tracing the phononic degrees of freedom (DOF) out of
the full electron-phonon density matrix (DM) $\hat{\rho}(t)$, 
we obtain the electronic DM $\hat{\rho}_\mathrm{e}(t) = \mathrm{Tr}_\mathrm{p}(\hat{\rho}(t))$ which 
obeys the effective Liouville equation \cite{Horsfield2004}
\begin{equation}
\label{eq:rhoeom1}
\dot{\hat{\rho}}_\mathrm{e}(t) = \frac{1}{\mathrm{i} \hbar} \
[ \hat{H}_\mathrm{e} , \hat{\rho}_\mathrm{e}(t) ]  
- \frac{1}{\mathrm{i} \hbar} \ [ \hat{F} , \hat{\mu}(t) ]
\end{equation}
where $\hat{\mu} (t) = \mathrm{Tr}_\mathrm{p}(\hat{X}\hat{\rho} (t))$.
ECEID's goal is to produce a closed set of equations of motion (EOM) for
$\hat{\rho}_\mathrm{e}(t)$ and the mean phononic occupancies
$N(t) = \mathrm{Tr}(\hat{N} \hat{\rho}(t))$, 
where $\hat{N} = \hat{a}^{\dagger} \hat{a}$ 
and $\hat{a}^{\dagger}$ ($\hat{a}$) is the phonon creation 
(annihilation) operator.
We take the exact form of the full DM
\begin{equation}
\label{eq:rhofull}
\hat{\rho}(t)  = \mathrm{e}^{- \frac{\mathrm{i}}{\hbar} \hat{H}_0 t}  \hat{\rho}(0) \mathrm{e}^{ \frac{\mathrm{i}}{\hbar} \hat{H}_0 t}  
-  \frac{1}{\mathrm{i} \hbar} \int_0^t 
\mathrm{e}^{\frac{\mathrm{i}}{\hbar} \hat{H}_0 (\tau-t)}  [ \hat{F} \hat{X}  , \hat{\rho}(\tau) ]
\mathrm{e}^{-\frac{\mathrm{i}}{\hbar} \hat{H}_0 (\tau-t)} \mathrm{d}\tau,
\end{equation}
insert it into the definition of $\hat{\mu}(t)$ and invoke two approximations.

First, we decouple the full DM into
$\hat{\rho}(\tau) \approx \hat{\rho}_\mathrm{e}(\tau) 
\hat{\rho}_\mathrm{p} (\tau)$, 
preserving electron-phonon correlation at lowest order in $\hat{F}$.
Second, we retain only single-phonon processes by ignoring terms of the form 
$\mathrm{Tr}_\mathrm{p}(\hat{a} \hat{a} \hat{\rho}(t) )$, $\mathrm{Tr}_\mathrm{p}(\hat{a}^\dagger 
\hat{a}^\dagger \hat{\rho}(t) )$.
This leads to the following equations \cite{Rizzi2016}
\begin{equation}
\hat{\mu}(t) = \frac{1}{M \omega} ( \mathrm{i} \ \hat{C}_\mathrm{c}(t) - \hat{A}_\mathrm{s}(t) ).
\label{eq:mueom1}
\end{equation}
and
\begin{equation}
\dot{N}(t) = \frac{1}{M \hbar \omega} \
\Big( \mathrm{i} \, \mathrm{Tr}_\mathrm{e}(\hat{F} \hat{C}_\mathrm{s}(t)) 
+ \mathrm{Tr}_\mathrm{e}(\hat{F} \hat{A}_\mathrm{c}(t)) \Big),
\label{eq:Neom1}
\end{equation}
where the auxiliary force operators $(\hat C_\mathrm c, \hat A_\mathrm c, \hat C_\mathrm s, \hat A_\mathrm s)$ obey the EOM
\begin{eqnarray}
\label{eq:dotcc}											
\dot{\hat{C}}_\mathrm{c} (t) &=& - 
\frac{\mathrm{i}}{\hbar} \ [ \hat{H}_{\mathrm{e}},\hat{C}_\mathrm{c}(t) ] 
+ \omega \hat{C}_\mathrm{s}(t) 
+ (N(t) + \frac{1}{2}) [\hat{F},\hat{\rho}_\mathrm{e}(t) ] \\
\dot{\hat{C}}_\mathrm{s} (t) &=& -\frac{\mathrm{i}}{\hbar} \ 
[ \hat{H}_{\mathrm{e}} , \hat{C}_\mathrm{s}(t) ] - \omega \hat{C}_\mathrm{c}(t) \\
\label{eq:dotac}	
\dot{\hat{A}}_\mathrm{c} (t) &=& 
- \frac{\mathrm{i}}{\hbar} \
[ \hat{H}_{\mathrm{e}}, \hat{A}_\mathrm{c}(t) ] 
+ \omega \hat{A}_\mathrm{s}(t)
+ \frac{1}{2} \ \{\hat{F}, \hat{\rho}_\mathrm{e}(t) \} \\
\label{eq:dotas}	
\dot{\hat{A}}_\mathrm{s} (t) &=& -\frac{\mathrm{i}}{\hbar} \
[ \hat{H}_{\mathrm{e}}, \hat{A}_\mathrm{s}(t) ] - \omega \hat{A}_\mathrm{c}(t).
\end{eqnarray}

Equations (\ref{eq:rhoeom1}, \ref{eq:mueom1}-\ref{eq:dotas}) constitute the ECEID set of EOM.
In this form, they still involve many-body electronic operators.
They can be further reduced to one-electron form by tracing out electrons, as described in \cite{Rizzi2016}. 

\subsection*{Open Boundaries in ECEID}
The derivation above describes a closed system. However electron injection and extraction requires opening the system to external reservoirs. We do this by the multiple probes OB method derived in \cite{McEniry2007}.

The system now consists of a central region connected to a left and a right lead. The central region contains the phonon DOF and is where the dynamical scattering and the electron-phonon energy exchange occur. The leads are usually metallic and some of their sites are connected to external probes, acting as particle baths. This setup effectively screens the finite size of the leads by broadening their discrete levels into a continuous spectrum.
In the wide band limit in the external baths \cite{McEniry2007}, the system's embedding self energy is
$\hat{\Sigma}^\pm = \mp(\mathrm{i} \Gamma/2) (\hat{I}_{\mathrm{L}} + \hat{I}_{\mathrm{R}} )$ where $\Gamma$ sets the coupling to the baths
and $\hat{I}_{\mathrm{L(R)}}$ are the identity operators over the regions coupled to the baths. We are using an orthonormal real space basis throughout.
The introduction of the OB transforms Eq. (\ref{eq:rhoeom1}) into \cite{McEniry2007}
\begin{equation}
i \hbar \dot{\hat{\rho}}_\mathrm{e}(t) = [ \hat{H}_\mathrm{e} , \hat{\rho}_\mathrm{e}(t) ]  - [ \hat{F} , \hat{\mu}(t) ] + \underbrace{\hat{\Sigma}^+ \hat{\rho}_\mathrm{e}(t) - \hat{\rho}_\mathrm{e}(t) \hat{\Sigma}^-}_{\mathrm{extraction}} 
+ \underbrace{\int_{-\infty}^{\infty} \Big( \hat{\Sigma}^{<}(E) \hat{G}^-_{\mathrm{S}}(E) - \hat{G}^+_{\mathrm{S}}(E) \hat{\Sigma}^{<}(E) \Big) dE}_{\mathrm{injection}}
\label{eq:eomob}
\end{equation}
where $\hat{\Sigma}^+ \hat{\rho}_\mathrm{e}(t) - \hat{\rho}_\mathrm{e}(t) \hat{\Sigma}^-$ represents electron extraction and the last term portrays electron injection.
The injection integral involves the retarded (advanced) Green's function for the lead-sample-lead system in the presence of the baths $\hat{G}^{+(-)}_{\mathrm{S}}(E)$ and the
quantity  $\hat{\Sigma}^{<}(E) = \frac{\Gamma}{2 \pi} ( f_\mathrm{L}(E) \hat{I}_\mathrm{L} +  f_\mathrm{R}(E) \hat{I}_\mathrm{R})$, where $f_\mathrm{L(R)}(E)$ are the desired incoming electronic distributions. 
These could describe a conventional applied electrochemical bias or, as we will see later, also electron beams targeted at specific energies.

The reduced one-electron form of the earlier ECEID EOM, discussed in \cite{Rizzi2016}, includes also a decoherence factor into the EOM of the auxiliary operators. 
Analogously	to the OB's role, that factor physically describes the same embedding of the finite system into an environment. For simplicity and physical consistency, we use the same decoherence strength, set by $\Gamma$, in both places.


\section*{A Simple Water Model}
\begin{figure}[h!tb]
	\includegraphics[width=1.0\columnwidth]{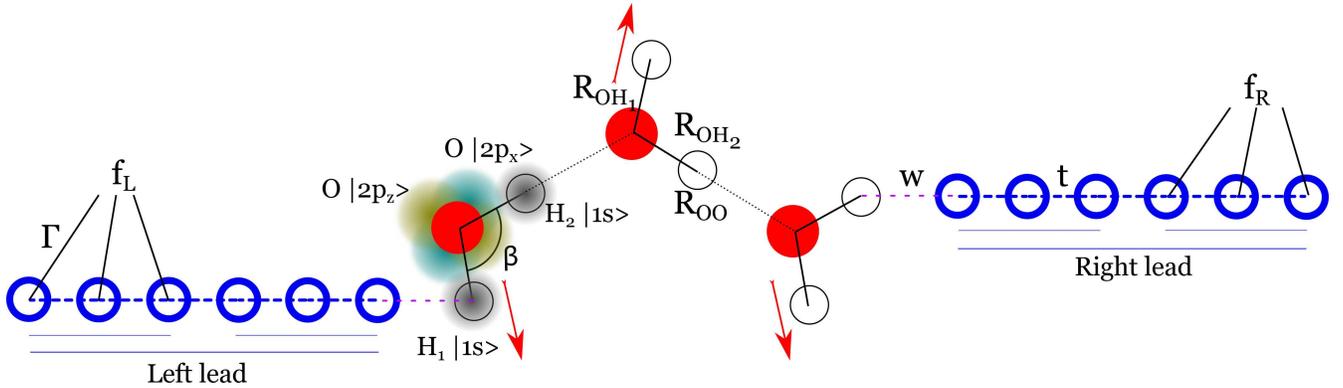}
	\caption{\label{fig:chainh2o}
		Reduced schematic of our model system, with a water chain of 3 molecules connected to its left and its right to metallic leads made of 6 atoms, 3 of which attached to external probes. See text for the dimensions of the system in the actual simulations. Each oxygen atom (red) has a $2p_{x}$ and a $2p_{z}$ orbital, while the hydrogen atoms (white) $H_{1}$ (pointing out of the chain) and $H_{2}$ (pointing in the chain) have a $1s$ orbital.
		The red arrows depict the phonon mode included in the calculations, with the $R_{OH_{1}}$ bonds vibrating in phase.
	}
	\centering
\end{figure}

To provide a qualitative study of the inelastic interaction of water with incoming electrons, we adopt a minimal 4-orbital tight-binding model for each water molecule.
We consider water molecules lying in the $xz$ plane and include a $1s$ orbital on each hydrogen atom $H_{1,2}$ and $2p_{x,z}$ orbitals on the oxygen atom $O$, with onsite energies $E_{H_{1,2}} = -13.61$ eV and $E_{O_{p_{x,z}}} = -14.13$ eV \cite{Paxton2011}. We place the Fermi energy $E_F$ halfway between the hydrogen and oxygen onsite energies and from now on we will use it as the zero of energy.

We connect water molecules to form a planar chain, whose equilibrium geometry was determined with DFT simulations in CP2K \cite{VandeVondele2005103} and is shown in Fig. \ref{fig:chainh2o}.  
The resulting electronic structure from DFT features a HOMO-LUMO gap of $6$ eV followed by a $1.6$ eV-wide group of unoccupied states. These are separated from the continuum by a $3$ eV gap, which probably originates from the 1-dimensional geometry of the chain. We set as our aim to reproduce the presence of the isolated lowest unoccupied band in our tight-binding model and to use it for the electron injection in the examples.
The Hamiltonian for the $j$th water molecule in the chain reads 
\[
\hat{H}_j = 
\begin{bmatrix}
E_{H_1}-E_F & \mp U_1 \cos\theta& U_1 \sin\theta & 0 \\
\mp U_1 \cos\theta & E_{O_{p_x}}-E_F & 0 & \pm U_2 \cos\theta \\
U_1 \sin\theta & 0 & E_{O_{p_z}}-E_F & U_2 \sin\theta \\
0 & \pm U_2 \cos\theta& U_2\sin\theta &E_{H_2}-E_F \\
\end{bmatrix}
\]
where upper signs match odd $j$s and lower signs even $j$s, $U_{1,2}$ are the hoppings between the oxygen and the hydrogen atoms $H_{1,2}$, $\beta = 105.8^{\circ}$ is the H-O-H angle and $\theta = (\pi - \beta)/2$. 
$H_1$ points out of the chain with a bond length $R_{OH_{1}} = 0.97 \ \text{\normalfont\AA}$ and $H_2$ forms the chain's backbone with $R_{OH_{2}} = 1.00 \ \text{\normalfont\AA}$
We choose a O-H hopping
\begin{equation}
U(R) = V \left(\frac{R_0}{R}\right)^2 \exp\left( 2 \left( - \left(\frac{R}{R_C}\right)^4 + \left(\frac{R_0}{R_C}\right)^4 \right) \right)
\end{equation}
where $V = 1.84 \, \hbar^2 / (2 m_e R_{0}^2)$, $m_e$ is the electron mass, $R_0 = R_{OH_{1}}$ is the equilibrium bond length and the critical length is set to $R_C = 1.8 \, R_{OH_{1}}$ \cite{Goodwin1989,Paxton2011}.
This makes $U_{1} = U(R_{OH_{1}}) = 7.45$ eV and $U_{2} = U(R_{OH_{2}}) = 6.84$ eV. 
The distance between the oxygen atoms is $R_{OO} = 2.67 \ \text{\normalfont\AA}$, so the inter-molecular hopping between $O \, 2p_{z}$ and $H_2$ is $U(R_{OO} - R_{OH_{2}}) = 0.57$ eV, while the hopping between the $O \, 2p_{x}$ orbital and $H_2$ is zero by construction.
This form of hopping has the advantage of recovering the result from Harrison's solid state table \cite{Harrison} $U(R_0) = V$ for equilibrium bond lengths, while decaying exponentially with large distances, suppressing intermolecular O-H interactions.

A DFT vibrational structure calculation for a 10 molecule water chain with periodic boundary conditions produces a set of phonon bands. One of them consists of phonon modes that predominantly involve out of chain hydrogens $H_1$ moving in and out of the chain. The modes in that band present closely clustered frequencies lying within a fraction of a percent of each other. For our present calculations, we are going to consider a fictitious mode (since the actual modes and their phases will of course depend on the specific boundary conditions) in which all $H_1$s are vibrating in phase with each other, as indicated in Fig. \ref{fig:chainh2o}. 
We considered different choices of relative phases for the vibrating $H$ atoms, but these did not produce a qualitative change in the results.
We take a representative value for the frequency $\hbar \omega = 0.473$ eV and for the vibrational reduced mass $M = 1$ a.m.u., for that band of modes.
A block of $\hat{F}$ for the $j$th water molecule in the chain reads
\[
\hat{F}_j = F
\begin{bmatrix}
0 & \mp \cos\theta & \sin\theta & 0 \\
\mp \cos\theta & 0 & 0 & 0 \\
\sin\theta & 0 & 0 & 0 \\
0 & 0 & 0 & 0 
\end{bmatrix}
\]
where $F = C \frac{\partial U(R)}{\partial R}\Bigr|_{R =
R_{OH_{1}}} = 6.60$ eV/$\text{\normalfont\AA}$, 
the upper signs correspond to an odd $j$, the lower ones to an even $j$ and the factor $C = 1 / \sqrt{10}$ arises because our phonon involves the motion of all the 10 water molecules in the chain.
The zero point amplitude of the phonon is $\sqrt{\braket{X^2}}_{N=0} = 0.064 \,
\text{\normalfont\AA}$, while the root mean square displacement of the
oscillator at $N=2$ is $\sqrt{\braket{X^2}}_{N=2} = 0.144 \,
\text{\normalfont\AA}$.

With the above value of $F$, in some simulations we observed violent oscillations
in the current that prevented the formation of a steady state in the system. We
believe that these oscillations are due to a limitation of the method
approximations in treating coherences for large $F$.
We can determine a critical $F$ by using a qualitative condition that compares a
typical electron-phonon transition rate with the inverse oscillator period 
$\frac{2\pi}{\hbar} F_{\textrm{c}}^2 X^2 D \simeq \frac{\omega}{2 \pi}$.
If we pick as a typical phonon displacement $X \simeq \sqrt{\braket{X^2}}_{N=0}$ and
a density of states $D \simeq \frac{1}{B}$ with $B \simeq 16$ eV the total bandwidth of the
water chain, we obtain $F_{\textrm{c}} \simeq 7$ eV/$\text{\normalfont\AA}$.
System dynamics with values about and above $F_{\textrm{c}}$ would involve
higher-order processes and coherences that ECEID cannot handle. 
The value above is indeed close to $F_{\textrm{c}}$. 
As a cure, consistent with the approximations in the method, we halve $F$ to $3.30$
eV/$\text{\normalfont\AA}$. 
The electron-phonon transition rates scale, to lowest order, quadratically with $F$; therefore, the above change would reduce the inelastic transition rates by a factor of $4$. Nevertheless, the underlying physics of the phenomenon that we are interested in, namely the inelastic electron injection in elastically forbidden energy ranges, remains unchanged.


\section*{Results and Discussion}
\begin{figure}[h!tb]
	\begin{center}
		\includegraphics[width=0.7\columnwidth]{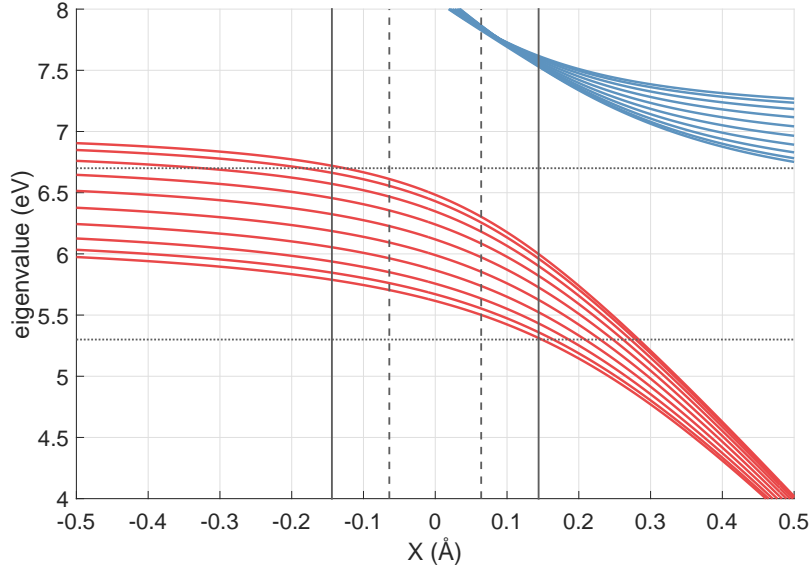}
		\caption{\label{fig:eigv10}
			Water chain eigenvalues of the levels above $E_F$ and their elastic variation with the phonon mode generalized coordinate $X$,
			with the FCB shown in red and the SCB in blue.
			The dashed vertical lines indicate the phononic zero point motion $\sqrt{\braket{X^2}}_{N=0}$ and the solid ones $\sqrt{\braket{X^2}}_{N=2}$. The horizontal dotted lines indicate the energies $E = 5.3$ eV and $E = 6.7$ eV that will be used in the electron-gun injection section.
		}
	\end{center}	
\end{figure}

We work with a chain made of $10$ water molecules.
Its electronic eigenvalues form $4$ bands that are symmetric with respect to $E_F$. 
The water chain eigenstates are initially populated at their ground state with a half filled band. 
The bandgap between the valence band and the first conduction band (FCB) is $11.2$ eV, while the gap between the FCB and the band above it is $1.6$ eV.
DFT calculations of the excited states of a corresponding system show the formation of a band analogous to the FCB and, above it, a continuum of states which is not captured in our simple TB model because of the reduced number of basis functions for water empty states. 
This work focuses on injecting electrons in the FCB and evaluating the inelastic contribution of the phonon. Effects involving the artificial second conduction band (SCB) will be commented on later.

To examine the effect of a classical phonon on the eigenspectrum, we add $-\hat{F} X$ to the water chain Hamiltonian, where $X$ is a classical generalized coordinate. A scan in $X$ samples the elastic effect of a frozen phonon.
In Fig. \ref{fig:eigv10} we show the variation with $X$ of the eigenvalues in the FCB. A positive $X$ makes $R_{OH_1}$ extend, reducing the FCB bandwidth and narrowing the gap with the valence band. A negative $X$ makes the FCB shift slightly upwards in energy.

Next, we connect the water chain to metal leads, as sketched in Fig. \ref{fig:chainh2o}, so that $H_1$ on the left and $H_2$ on the right of the chain are linked to the neighbouring metal site by $w=-2$ eV. 
The metallic leads serve as a tool for injection and, initially, they are not populated.
They are made of 80 sites each, with a single state per site and a nearest neighbour hopping $t=-4$ eV. Their onsite energy is $10$ eV, so that the metallic energy band ranges from $2$ to $18$ eV, overlapping only with the water's conduction bands. 
The $40$ leftmost sites in the left lead and the $40$ rightmost ones in the right lead are coupled by $\Gamma = 1.25$ eV to external probes for the OB.

\subsection*{Electron-Pulse Injection}
\begin{figure}[h!tb]
	\begin{center}
		\includegraphics[width=1.0\columnwidth]{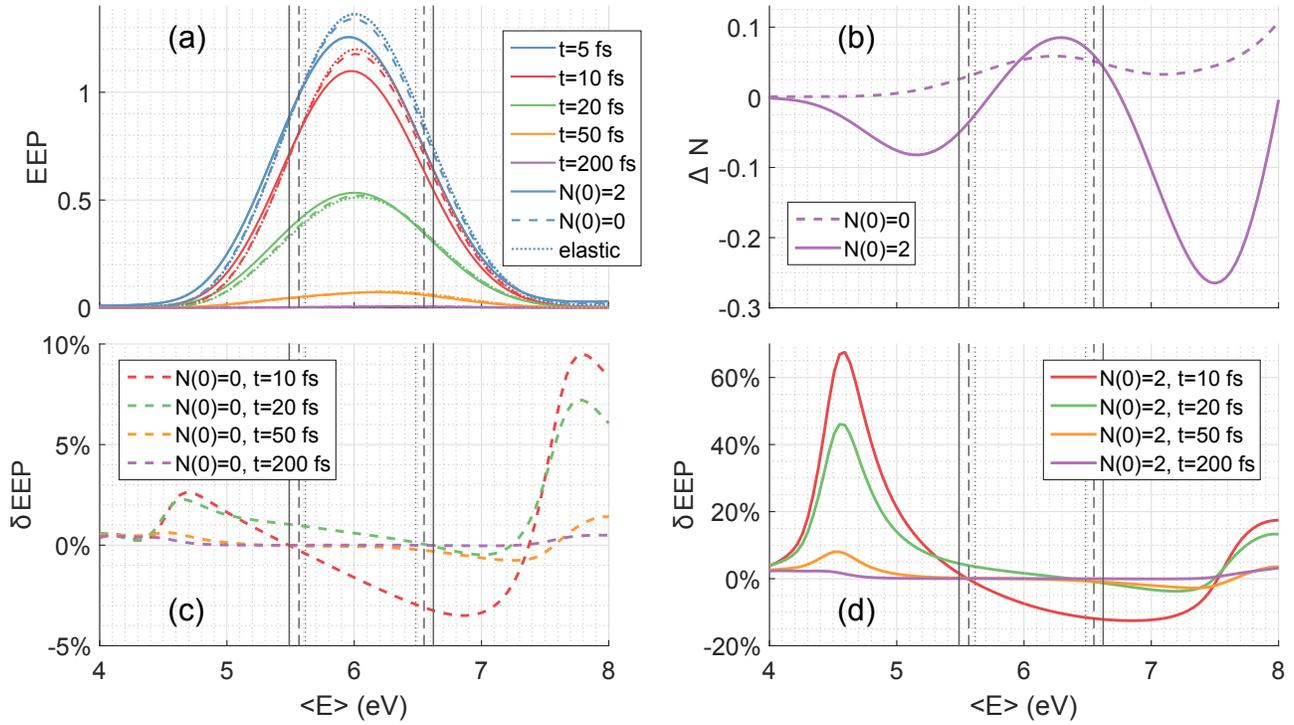}
		\caption{\label{fig:gaussexc}
			Electron-pulse simulations where Gaussian wavepackets are injected into the water chain. The vertical dotted lines mark the edges of FCB for $X=0$; 
the dashed vertical lines mark the range spanned by FCB for $|X| < \sqrt{\braket{X^2}}_{N=0}$; the solid vertical lines mark this range for $|X| < \sqrt{\braket{X^2}}_{N=2}$.
			Panel (a): FCB excess electron population for wavepackets with an average energy $\braket{E}$ at a range of times.
			Solid lines correspond to the phonon starting at $N(0) = 2$, dashed lines to $N(0) = 0$ and dotted lines to the elastic case.
			Panel (b): phonon occupancy variation $\Delta N$ at $t = 200$ fs for initial $N(0) = 0$ (dashed) and $N(0) = 2$ (solid).
			Panel (c): difference between the inelastic excess electron population with $N(0) = 0$ and the elastic one at different times, normalized by the elastic case at $t = 5$ fs.
			Panel(d): same as panel (c), with the inelastic case of $N(0) = 2$.
		}
	\end{center}	
\end{figure}
We begin with probably the most intuitive picture of electron injection, namely firing actual electronic wavepackets from the left lead towards the water chain and propagating the ECEID EOM.
Initially, we introduce in the empty leads an electron pulse in the form of a spin-degenerate wavepacket moving from the left lead towards the water chain. 
The pulse has the following form 
\begin{equation}
\ket{\Psi_0}  = C \sum_n \mathrm{e}^{-(n-n_0)^2 / 2 \sigma^2} \mathrm{e}^{\mathrm{i} k n} \ket{n}
\end{equation}
where $n$ spans the left lead atomic basis, $n_0 = 55$ is the pulse's central site, $\sigma = 12$ sites, $k \in [0,\pi]$ is a dimensionless crystal momentum and $C$ is the normalization factor. By scanning over $k$, we can vary the energy of the incident wavepacket which has an initial full width at half maximum in energy of about $1$ eV.
We keep only the extraction term in the OB in Eq. (\ref{eq:eomob}), so that the backscattered or transmitted parts of the wavepacket reaching the ends of the leads can get absorbed and not reflected back into the water.

To measure electron absorption into the water, we sum over the occupations of the FCB. We call this quantity excess electron population (EEP) and show it at different times in Fig. \ref{fig:gaussexc}(a). 
We perform a phonon-free elastic simulation (dotted line) and inelastic simulations where the phonon is initialized at $N(0) = 0$ (dashed line) or $N(0) = 2$ (solid line). 
At $t = 5$ fs the EEP displays a peak for an incident energy in the middle of the FCB. 
It also displays electron absorption for incident energies well outside the FCB. 
The inelastic contribution to injection in this elastically forbidden range is intertwined with elastic effects due to the energy width of the pulse.
In fact, the elastic curves are not significantly different from the inelastic ones at this level of analysis. With time progressing, the EEP decreases in magnitude as electrons leak back out into the leads, until, at $t=200$ fs, the EEP drops close to zero in all cases. 

Nevertheless, the electron pulse leaves a long lasting inelastic mark on the phonon occupancy variation $\Delta N (t) = N(t) - N(0)$, shown at $t=200$ fs in Fig. \ref{fig:gaussexc}(b). $\Delta N (t)$ displays a radically different behaviour depending on the initial condition $N(0)$ and on the pulse average energy $\braket{E}$.
Electron injection in the elastically forbidden range, outside the FCB, hinges on the phonon, as the electron-phonon interaction enables electrons to access the water states via phonon emission or absorption. This inelastically-assisted electron injection furthermore is controlled by the effective vibrational temperature, as the value of $N(0)$ determines which phonon-assisted processes are allowed. 

For pulses below the water FCB edge, $\braket{E} < 5.6$ eV, an electron must absorb phonons to enter the water; therefore the inelastic hopping can only be activated if $N(0) > 0$. Indeed, we see that the dashed curve representing $N(0) = 0$ in Fig. \ref{fig:gaussexc}(b) remains close to zero at low energies and starts increasing for pulses with $\braket{E} \simeq 5$ eV. At that energy range, the high energy components of the pulse can enter the FCB elastically and trigger further inelastic processes.
On the other hand, the solid curve, for $N(0) = 2$, shows a clear dip in $\Delta N$
at incident energies about $\hbar\omega$ below FCB, highlighting electron injection assisted by phonon absorption.

For incident pulses above the upper band edge, $\braket{E} > 6.5$ eV, electrons have to emit a phonon to enter the water FCB, so one expects a $\Delta N > 0$. 
However, the higher energy SCB (see Fig. \ref{fig:eigv10}) obscures the phonon emission process in this range by mixing it with phonon absorption, leading to injection into SCB.

To isolate the inelastic contribution in the EEP, we introduce $\delta$EEP: the difference between the inelastic EEP and the elastic EEP, normalized by the elastic EEP at 
$t = 5$ fs. This quantity measures the effective inelastic contribution to injection. 
In Fig. \ref{fig:gaussexc}(c-d), we show $\delta$EEP for $N(0) = 0$ and $N(0) = 2$ respectively. 
The latter shows an intense peak at $\braket{E} \simeq 4.5$ eV, corresponding to
phonon absorption by the high-energy components of the incident wavepacket.
This signature peak is much weaker in the $N(0) = 0$ case.
Thus $N$, and the effective phonon temperature, is a key controlling factor for the intensity of phonon-assisted injection.
Above the FCB, we see another increase in $\delta$EEP marking inelastic injection in the FCB, but mixed with the concomitant injection in the SCB.

The decay time of the injected electrons shows a dependency on $\braket{E}$, with $\delta$EEP for pulse energies outside the FCB decaying slower than for energies inside the FCB. This behaviour suggests an energy dependent lifetime of the FCB states, which we will explore more thoroughly later.

In the electron-pulse simulations, the energy spread of the wavepackets obscures the inelastic effects on the FCB by mixing them with elastic contributions and the injection in the SCB. To overcome these limitations, and probe the inelastic injection mechanism further, we now exploit the OB to send in a steady incident electron beam with a sharp energy spectrum.

\subsection*{Electron-Gun Injection}
\begin{figure}[h!tb]
	\begin{center}
		\includegraphics[width=0.8\columnwidth]{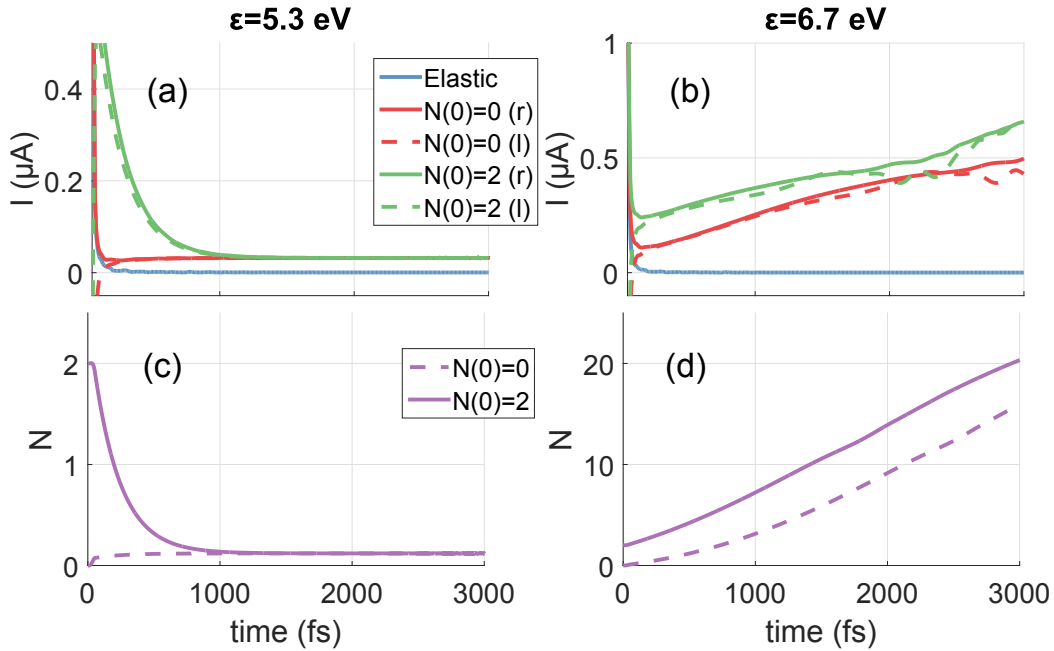}		
		\caption{\label{fig:egunsum}
			Electron-gun simulations where electrons are injected within a narrow energy window, just above or below the water FCB. Panels (a) and (b): current measured in the second metal bond on the right (solid) and on the left (dashed) of the chain with the electron gun aimed at $E = 5.3 \pm 0.1$ eV and $E = 6.7 \pm 0.1$ eV, respectively. Elastic ECEID simulations are in blue, $N(0) = 0$ in red and $N(0) = 2$ in green.
			Panels (c) and (d): dynamics of $N$ corresponding to (a) and (b), with $N(0) = 0$ (dashed) and $N(0) = 2$ (solid).
		}
	\end{center}	
\end{figure}
We make use of the electronic OB equation (\ref{eq:eomob}) to inject a steady electron beam, as opposed to the single wavepacket above.
Here the leads are coupled to baths kept at zero electronic temperature and zero electrochemical potential. Thus $f_\mathrm{L/R}(E)$ are the corresponding equilibrium electronic distributions.
However, in addition, $f_\mathrm{L}(E)$ contains a top-hat spike between $E= \epsilon - \delta \epsilon$ and $E = \epsilon + \delta \epsilon$. This setup provides a steady electron flux hitting the water chain.
We concentrate on the interesting scenario found earlier when the incoming electron energies are just above or below the water FCB, so that phonon-activated electron injection is the dominant process.
We aim the electron gun at $\epsilon = 5.3$ eV with $\delta \epsilon = 0.1$ eV, just below the chain FCB $5.6 < E < 6.5$ eV and just above it at $\epsilon = 6.7$ eV, as can be seen from the horizontal dotted lines in Fig. \ref{fig:eigv10}.

In Fig. \ref{fig:egunsum}(a-b) we show the current measured on the right and on the left of the water chain in ECEID simulations, with the initial phonon occupancies $N(0)=0$, $N(0)=2$ considered earlier, together with a purely elastic, phonon-free simulation. 
In all cases, the injection energy is out of the chain's elastic transmission range, so the elastic current remains very close to zero as expected.
At the lower injection energy $\epsilon = 5.3$ eV, if $N(0) > 0$, electrons can hop into the water FCB through phonon absorption. 
Indeed, at $N(0)=2$, the current decreases in time, with the phonon cooling down, as shown in Fig. \ref{fig:egunsum}(c). This current-assisted cooling may look counterintuitive if one imagines a current to always heat its surroundings, but in fact current-assisted cooling in molecular systems is not unfamiliar \cite{McEniry2009}. Here the phonon cooling and the nonzero current are a signature of the phonon absorption injection mechanism.
 
At $N(0)=0$ and $\epsilon = 5.3$ eV, even if no injection is expected, there is a small increase in current accompanied by an increase in $N$, until both the current and $N$ eventually reach a steady state matching the one for $N(0)=2$. This is caused by the broadening of the FCB due to the presence of the leads, together
with the fact that the electron-gun window of populated states in the leads effectively
constitutes an electronic excitation above the ground state of the system. Therefore, thermodynamically, a tendency of the phonon to equilibrate at a raised energy above the vibrational ground state is to be expected. The broadening of the FCB density of states - and the small heating effect starting from $N(0)=0$ - decrease with decreasing water-metal coupling ($w$).

In the $\epsilon = 5.3$ eV results, the currents evaluated on the left and on the right of the chain are superimposable: the system dynamics can be approximated by a series of steady states. This does not happen for $\epsilon = 6.7$ eV, where left and right currents tend to slightly differ at all times, indicating that the system dynamics does not manage to remain in an electronic steady state during the heating of the phonon. The mechanism at play now is electron injection into water through phonon emission, with a positive feedback mechanism. The incoming electron flux causes phonon heating through emission, and a resultant current due to the electrons inelastically injected in the water.
The increase in $N$, in turn, increases the electron-phonon scattering rate (from the Golden Rule), further increasing the inelastic electron current.
In the present model, the water is not coupled to a vibrational thermal bath and this leads to the dramatic phonon heating see in Fig. \ref{fig:egunsum}(d). 

Of course, the vibrational energy cannot increase without bounds, but the principle remains: a potentially powerful injection mechanism from elastically forbidden energies, with the vibrational effective temperature as a key controlling factor. From a simulation point of view, the absence of a vibrational thermostat (sometimes referred as the completely undamped limit in the context of inelastic nanoscale transport \cite{Frederiksen2007}) offers the advantage that a single real-time ECEID simulation, with $N$ allowed to respond and scan a range of values, probes a range of different temperature-dependent regimes. 

In reality, there are two limiting mechanisms at play. On the one hand, the phonon mode will be coupled to the environment, typically through hydrogen-bonding to other water molecules. Therefore, part of this energy will be used in heating the aqueous environment. This energy transfer will correspond to a specific time scale related to the thermal conductivity of water, which is relatively low. If energy is pumped into this phonon mode at a faster rate, then it will accumulate in the bond. Here the harmonic approximation breaks down, and the inclusion of anharmonic potentials that incorporate the possibility of bond dissociation becomes essential. Such a resonant damage mechanism is reminiscent of dissociative electron attachment in DNA-related systems \cite{Haxton2004,Smyth2014}, where electrons trapped in a resonant state transfer sufficient energy to a vibrational mode to break the corresponding bond. Incoming electrons with a well defined energy could trigger inelastic effects such as the ones described in the present paper and be a prominent cause for dramatic heating and bond breaking.

\subsection*{Eigenstate Lifetime and Bandedge Trapping}
\begin{figure}[h!tb]
	\begin{center}
		\includegraphics[width=0.7\columnwidth]{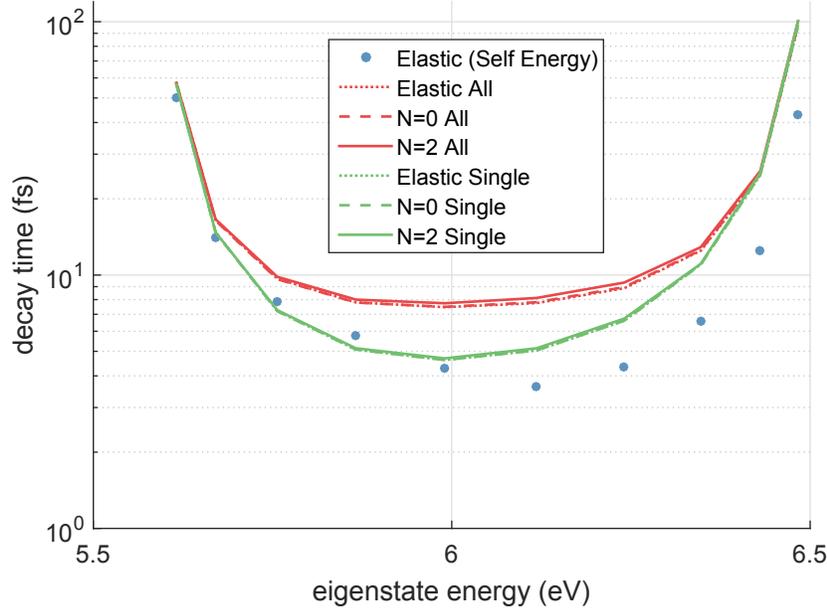}		
		\caption{\label{fig:decaycomp}
			Lifetime comparison of the water FCB eigenstates for: static self energy calculations ($\tau_{\Sigma,j}$) (solid dots), exponential fits of the decaying population of a single initially populated level ($\tau_{s,j}$) (green curves), exponential fits of the total population of all excited states, starting from a single initial populated FCB level ($\tau_{a,j}$) (red curves). Dotted curves correspond to elastic ECEID real-time calculations, dashed and solid lines to $N(0) = 0$ and $N(0) = 2$ ECEID simulations, both with $\dot{N}$ kept equal to $0$.
		}
	\end{center}	
\end{figure}
In Fig. \ref{fig:gaussexc}(c-d) we noticed that the $\delta$EEP decays slower for pulses above and below the FCB. This observation provides an indirect measure for the lifetime of the inelastically injected electrons in the water as a function of the initial injection energy. 
We know that electrons entering the water from these elastically forbidden energies do so by phonon emission and absorption, landing in the highest and lowest energy states in the FCB. If these states were long lived, they would act as an effective electron trap. 

To understand this phenomenon, we evaluate the lifetimes of all eigenstates, $j$, in the water FCB, calculated in two different ways. The first is the purely elastic lifetime against tunnelling out into the leads given by
$\tau^{-1}_{\Sigma,j} = - \frac{2}{\hbar} \textrm{Im} \braket{j | \hat{\Sigma}^+_{\mathrm{c}}(E_j) |j}$
where $\hat{\Sigma}^+_{\mathrm{c}}$ is the water chain's self-energy which incorporates the embedding in the environment through the end sites of the water chain that are coupled to the leads.
The variations in $\tau^{-1}_{\Sigma,j}$, over the given narrow energy range, come mainly from the amplitudes of the water states $\ket{j}$ on the end sites of the chain.
The second is directly from dynamical simulations, in which we release an excess electron from a chosen water eigenstate, propagate the system in time and fit an exponential decay to the decreasing EEP.
We perform ECEID calculations for the purely elastic case and for the two earlier phonon occupancies $N(0)=0$ and $N(0)=2$. 
We include only the OB extraction term and, to concentrate on the lifetime, we keep $N$ frozen at its initial value. 
In the real-time simulations we further calculate and compare the lifetime associated with the population of the chosen initial state itself $\tau_{s,j}$ and the sum of populations of all FCB states $\tau_{a,j}$.

The characteristic horseshoe shape that all the decay times display in Fig. \ref{fig:decaycomp} fundamentally originates from the dominant elastic mechanism. It can be understood from the form of the self-energy and the chain eigenstates. More generally, of course, the lifetime against escaping into the environment will depend on the details of the system-environment coupling. However, as our specific example illustrates, one may in general expect different system excited states to have significantly different lifetimes against the environment.
The self-energy lifetime $\tau_{\Sigma,j}$ tends to agree with $\tau_{s,j}$, especially in the low energy range, while $\tau_{a,j}$ is longer throughout the energy range.
Any difference between $\tau_{s,j}$ and $\tau_{a,j}$ signals higher-order scattering processes, not captured by the perturbative $\tau_{\Sigma,j}$.
Elastic ECEID calculations agree closely with the inelastic ones, indicating that the leading electron escape mechanism from the FCB is the elastic one to the leads, similarly to what was observed in 
earlier, perturbative inelastic tunnelling simulations \cite{Galperin2001}. The lifetimes computed here are comparable to the resonance lifetimes reported for tunnelling through 3d water structures in \cite{Peskin1999,Galperin2001a}.

The states at the band edge have a lifetime about one order of magnitude longer than the rest. They are the states most involved in the electron-phonon injection discussed earlier. 
A focused phonon-assisted electron injection in combination with the energy-dependent lifetimes of the FCB states, can thus provide an effective trapping mechanism for incident carriers, dependent on their incoming energy.

\section*{Conclusions}

We have performed dynamical non-adiabatic non-equilibrium electron-phonon quantum simulations of the inelastic injection and subsequent dynamics of excited electrons in a model water system. Such a real-time quantum approach will always be among the more computationally challenging methods for this class of problems, but it is essential for understanding the microscopic processes involved. 
We have seen that electron-phonon interaction provides a key mechanism by which excited electrons can be generated in water, as inelastic electron-phonon injection is a critical process in granting incoming electrons access to the water excited states.
In addition, we have seen that the vibrational temperature is a crucial controlling factor in the presence of an incoming electron flux. Depending on the incoming electron energy, the electron-phonon interaction can activate a current-assisted cooling or an exponential heating, where the dynamics of the system can further deviate from a steady-state description. 
By exploiting the energy-dependent lifetime of the water chain states, the injection of electrons in a specific energy ranges can result in partial electron trapping in the water.
It is hoped that these insights into the dynamical interaction of low energy electrons and water will lead to further more realistic simulations probing cell damage mechanisms and to the development of improved models of radiation exposure.
	
\bibliography{PAPERS,extra}

\section*{Acknowledgements}
We thank the Leverhulme Trust for funding this research under grant RPG-2012-583 and Alfredo Correa for useful discussions.

\section*{Author contributions statement}


V.R. performed the simulations. All authors analysed the results and reviewed the manuscript. 

\section*{Additional information}

The authors declare no competing financial interests.
	
\end{document}